\documentstyle[amsfonts,12pt]{article}
\normalbaselines
\input tcilatex
\QQQ{Language}{
American English
}

\begin{document}

\title{Hypersymmetry:\linebreak[4]a ${\Bbb Z}_3$-graded generalization of
Supersymmetry.}
\author{Viktor Abramov\thanks{%
Permanent Address: Institute of Pure Mathematics, Tartu University,
Vanemuise 46, Tartu, Estonia}\linebreak[4] \and Richard Kerner \and Bertrand
Le Roy \\
{Laboratoire de Gravitation et Cosmologie Relativistes}\\
{Universit\'e Pierre et Marie Curie - CNRS URA 769}\\
{Tour 22, 4-\`eme \'etage, Bo\^\i te 142}\\
{4, Place Jussieu, 75005 Paris, France}}
\maketitle

\begin{abstract}
We propose a generalization of non-commutative geometry and gauge theories
based on ternary ${\Bbb Z}_3$-graded structures. In the new algebraic
structures we define, we leave all products of two entities free, imposing
relations on ternary products only. These relations reflect the action of
the ${\Bbb Z}_3$-group, which may be either trivial, i.e. $abc=bca=cab$,
generalizing the usual {\it commutativity}, or non-trivial, i.e. $abc=jbca$,
with $j=e^{(2\pi i)/3}$. The usual ${\Bbb Z}_2$-graded structures such as
Grassmann, Lie and Clifford algebras are generalized to the ${\Bbb Z}_3$%
-graded case. Certain suggestions concerning the eventual use of these new
structures in physics of elementary particles are exposed.

Proposed PACS classification numbers:

03.65.F, 11.30.P, 12.40.-y, 14.65.-q\newpage 
\end{abstract}

\section{Introduction}

\TeXButton{baselineskip}{\baselineskip=24pt}In a recent series of articles 
\cite{cubdirac, cubsupsym, z3struct} we have advocated the ${\Bbb Z}_3$%
-grading as a natural generalization of well-known ${\Bbb Z}_2$-graded
structures, such as graded Lie algebras, superspaces and ${\Bbb Z}_2$-graded
generalizations of non-commutative geometry \cite{duboisncg,
duboissupermatgeo}. Most of the cases in which the ${\Bbb Z}_3$-grading was
studied has been based on the grading of ordinary algebras of matrices or
operators.

In this article we wish to stress the fact that the natural structure on
which the ${\Bbb Z}_3$-grading takes its full meaning is {\it a ternary
algebra}, which means a linear vector space over complex numbers on which a 
{\it ternary composition law} is defined. Although ternary laws can be
modelled in ordinary algebras with an associative binary law by defining
corresponding ternary ideals and dividing the algebra by the equivalence
relations induced by these ideals, one may also introduce ternary
composition laws for the entities which can not be derived from a binary law.

Although we believe that these novel algebraic constructions might be
pertinent for the description of quark fields and new models of elementary
interactions in particle physics, we shall stress here mathematical rather
than physical aspects, keeping hope that further developments and physical
applications of ternary structures will follow soon.

During the last decade a spectacular development of non-commutative
generalizations of differential geometry and Lie group theory has been
achieved; the respective new chapters of mathematical physics are known
under the names of {\it Non-Commutative Geometry} and {\it Quantum Groups
and Quantum Spaces}. In both cases, the crucial question asked concerned the
behavior of a ``product'', or in more general terms, of a {\it binary
composition law} under the permutation of two ``factors''. We shall
generalize this approach for the case in which no conditions are imposed on
binary products (which may even not be defined in some cases), but in
contrast, specific behavior of ternary composition laws will be required.

In what follows, we shall briefly recall the action of the permutation group
of three elements, $S_3$, in the complex plane, and apply its different
representations to ternary composition laws. Next, we shall generalize the
notions of Grassmann algebras, superspaces, supermatrices, Lie algebras and
Clifford algebras replacing systematically binary symmetry conditions by
their ternary counterparts.

Although many questions concerning the use of such ternary algebras in field
theory remain unanswered, and a lot of constructions have still to be
invented, we believe that this topic may attract attention of theoretical
physicists.

\section{The actions of ${\Bbb Z}_3$ and $S_3$ on ternary products.}

The group of permutations of three objects contains six elements, three of
which form the abelian subgroup ${\Bbb Z}_3$. These permutations are: 
\[
\left( 
\begin{array}{ccc}
A & B & C \\ 
A & B & C
\end{array}
\right) \left( 
\begin{array}{ccc}
A & B & C \\ 
B & C & A
\end{array}
\right) \left( 
\begin{array}{ccc}
A & B & C \\ 
C & A & B
\end{array}
\right) \left( 
\begin{array}{ccc}
A & B & C \\ 
C & B & A
\end{array}
\right) \left( 
\begin{array}{ccc}
A & B & C \\ 
B & A & C
\end{array}
\right) \left( 
\begin{array}{ccc}
A & B & C \\ 
A & C & B
\end{array}
\right) 
\]

The first three elements represent {\it cyclic} permutations, including the
identity. The entire group $S_3$ can be generated by two elements, a cyclic
permutation and one of the three involutions (odd permutations).

The important fact about the group $S_3$ is that it is the last of
permutation groups having a faithful representation in the complex plane;
the next permutation group, $S_4$, containing 24 elements, has a
representation in the complex plane that is not faithful, and starting from $%
S_5$, the permutation groups do not have representations in complex plane,
besides the trivial and reduced representation assigning an involution to
all odd elements and the identity to all even elements.

This fact is well known since Galois, and it is this very phenomenon that
makes possible the existence of algebraic solutions (in complex numbers) of 2%
$^{\text{nd}}$, 3$^{\text{rd}}$ and 4$^{\text{th}}$ order equations only,
excluding the algebraic solutions of any equation of order 5 and higher.

We shall represent the ${\Bbb Z}_3$ group in the complex plane with the
multiplications by the cubic roots of 1, i.e. $1,j$ and $j^2$, where 
\[
j=e^{2i\pi /3}\text{, }\ j^2=e^{-2i\pi /3}\text{,\ and\ }1+j+j^2=0 
\]
The odd permutations can be generated by complex conjugation; the remaining
two odd permutations, corresponding to reflections in the roots $j$ and $j^2$
can be obtained via composition of complex conjugation with one of the
cyclic elements (i.e. the rotations by $2\pi /3$ and $4\pi /3$, represented
as multiplications by $j$ and $j^2$ respectively).

In our generalizations of binary non-commutative structures we shall
systematically replace the representations of the group ${\Bbb Z}_2$ acting
on binary relations by the representations of the group ${\Bbb Z}_3$ acting
on ternary relations. Whenever these relations can be represented as induced
by an ordinary binary composition rule in some associative algebra, we shall
suppose that the binary products are totally independent.

Of course, in our generalization we cannot distinguish the representations
of ${\Bbb Z}_2$ from that of the group $S_2$, because the groups are
identical; here we have the choice between the cyclic group ${\Bbb Z}_3$ or
the whole permutation group $S_3$ which has six elements.\ The resulting
algebraic structures are very different too.

It is useful to remind that all binary relations can be interpreted in two
alternative ways, depending on whether we write them on one side of the
equation, or with non-trivial left- and right-hand sides; the ternary
generalizations will impose stronger or weaker conditions when interpreted
in these alternative manners.

Here are the examples of ternary generalizations of well-known binary
structures that we shall study one by one in the next sections. We shall
start with the classification of bi- (respectively, tri-linear) mappings
from vector spaces into complex numbers. The bilinear forms can be separated
into different categories following their ${\Bbb Z}_2$-representation
properties: 
\[
(X,Y)=(Y,X)\text{,\ a\ trivial\ representation\ of\ }{\Bbb Z}_2 
\]
which is called a symmetric 2-form. The same condition can be written as 
\[
(X,Y)+(-1)(Y,X)=0 
\]
which in binary case turns out to be equivalent with the former one.

In the case of ternary generalization (3-linear forms satisfying given
representation properties with respect to the group ${\Bbb Z}_3$) similar
conditions are no more equivalent: 
\[
(X,Y,Z)=(Y,Z,X)=(Z,X,Y)\text{,\ a\ trivial\ representation\ of\ }{\Bbb Z}_3%
\text{,} 
\]
the second interpretation will lead to the following condition: 
\[
(X,Y,Z)+j(Y,Z,X)+j^2(Z,X,Y)=0 
\]
Here the first condition implies the second one, whereas from the second
condition may follow either the first solution (i.e. all cyclic permutations
being equal), or a second (and the only other possible) one, namely 
\[
(X,Y,Z)=j^2(Y,Z,X)=j(Z,X,Y) 
\]
which is stronger and obviously satisfies the above equation.

A similar scheme can be applied for the action of the whole $S_3$ group if
we decide how to represent the odd permutations. If only the action of $%
{\Bbb Z}_3$ is represented, this means that the result of (Z,Y,X) is
independent from (X,Y,Z); if not,the complex conjugation in complex plane
provides us with a representation of $S_3$ when combined with the rules for $%
{\Bbb Z}_3$: 
\[
(X,Y,Z)=\overline{(Z,Y,X)} 
\]
On the other hand, if we decide that $(X,Y,Z)=(Z,Y,X)$, this will also
define an action of $S_3$, which in this case will not be faithful anymore.

The skew-symmetric bilinear forms are generalized by 
\[
(X,Y,Z)=j(Y,Z,X)=j^2(Z,X,Y) 
\]

or 
\[
(X,Y,Z)+(Y,Z,X)+(Z,X,Y)=0 
\]
in the case of a ${\Bbb Z}_3$-generalization or 
\[
(X,Y,Z)+(Y,Z,X)+(Z,X,Y)+(Z,Y,X)+(Y,X,Z)+(X,Z,Y)=0 
\]
in the case of the $S_3$-generalization.

\section{${\Bbb Z}_3$-graded analogue of Grassmann algebra}

\label{z3grass}Perhaps the simplest and the most straightforward ${\Bbb Z}_3$%
-graded generalization of a well-known binary algebraic structure is the
Grassmann algebra.

Consider a finite-generated associative free algebra over complex numbers.
Let us denote the generators of this algebra by $\theta ^A,\theta ^B$, $%
A,B=1,2,..,N$. We suppose that the $N^2$ products $\theta ^A\theta ^B$ are
linearly independent entities, whereas the products of three generators $%
\theta ^A\theta ^B\theta ^C$ are subjected to the following condition: 
\[
\theta ^A\theta ^B\theta ^C=j\theta ^B\theta ^C\theta ^A 
\]

The immediate corollary is that any product of four or more generators must
vanish. Here is the proof: 
\[
(\theta ^A\theta ^B\theta ^C)\theta ^D=j\theta ^B(\theta ^C\theta ^A\theta
^D)=j^2(\theta ^B\theta ^A\theta ^D)\theta ^C=\theta ^A(\theta ^D\theta
^B\theta ^C)=j\theta ^A\theta ^B\theta ^C\theta ^D, 
\]
Now, as $(1-j)\not =0$, one must have $\theta ^A\theta ^B\theta ^C\theta
^D=0 $. The dimension of the ${\Bbb Z}_3$-graded Grassmann algebra is ${%
N(N+1)(N+2)/3}+1$. Any cube of a generator is equal to zero; the odd
permutation of factors in a product of three leads to an independent
quantity.

Our algebra admits a natural ${\Bbb Z}_3$-grading: under multiplication, the
grades add up modulo 3; the numbers are grade 0, the generators $\theta ^A$
are grade 1; the binary products are grade 2, and the ternary products grade
0 again. The dimensions of the subsets of grade 0, 1 and 2 are,
respectively, $N$ for grade 1, $N^2$ for grade 2 and $(N^3-N)/3+1$ for grade
0.

The lack of symmetry between the grades 1 and 2 (corresponding to the
generator $j$ and its square $j^2$ in the cyclic group ${\Bbb Z}_3$, which
are interchangeable, suggests that one should introduce another set of $N$
generators of grade 2, whose squares would be of grade 1, and which should
obey the conjugate ternary relations as follows: 
\[
\bar \theta ^{\bar A}\bar \theta ^{\bar B}\bar \theta ^{\bar C}=j^2\bar
\theta ^{\bar B}\bar \theta ^{\bar C}\bar \theta ^{\bar A} 
\]
With respect to the ordinary generators $\theta ^A$, the conjugate ones
should behave like the products of two $\theta $'s, i.e. 
\begin{equation}
\theta ^A(\theta ^B\theta ^C)=j(\theta ^B\theta ^C)\theta ^A\rightarrow
\theta ^A\bar \theta ^{\bar B}=j\bar \theta ^{\bar B}\theta ^A
\label{z2grad0}
\end{equation}
and consequently, 
\begin{equation}
\bar \theta ^{\bar B}\theta ^A=j^2\theta ^A\bar \theta ^{\bar B}
\label{z2grad0bis}
\end{equation}
One may also note that there is an alternative choice for the commutation
relation between the ordinary and conjugate generators, that makes the
conjugate generators different from the binary products of ordinary
generators: 
\begin{equation}
\theta ^A\bar \theta ^{\bar B}=-j\bar \theta ^{\bar B}\theta ^A\text{ and }%
\bar \theta ^{\bar B}\theta ^A=-j^2\theta ^A\bar \theta ^{\bar B}
\label{z2grad1}
\end{equation}
which are still compatible with the ternary relations introduced above.

This could be interpreted in the following way. We have assumed that the
algebra's field is the field of complex numbers, but we can imagine that it
is possible to multiply an element of the ${\Bbb Z}_3$-graded Grassmann
algebra by an element of a {\it binary} Grassmann algebra. We assume that
the binary elements commute with the ternary ones, but anticommute as usual
with each other. The ${\Bbb Z}_3$-graded Grassmann elements of a given grade
still have no binary commutation relation. Then, our new algebra admits two
gradings: the ${\Bbb Z}_2$-grading and the ${\Bbb Z}_3$-grading. The
elements of ${\Bbb Z}_2$-grade 0 and ${\Bbb Z}_3$-grades 1 and 2 obey the
rules (\ref{z2grad0}) and (\ref{z2grad0bis}) whereas the elements of ${\Bbb Z%
}_2$-grade 1 and ${\Bbb Z}_3$-grades 1 and 2 obey the rules (\ref{z2grad1}).
If we think that these objects can help in modelling of the quark fields,
then a quark variable would be of ${\Bbb Z}_2$-grade 1 and ${\Bbb Z}_3$%
-grade 1, and an antiquark variable of ${\Bbb Z}_2$-grade 1 and ${\Bbb Z}_3$%
-grade 2. Then, the products of a quark and an antiquark would have both
grades zero, making it a boson. In the same way, the products of three quark
or three antiquark fields would be of ${\Bbb Z}_3$-grade 0 and of ${\Bbb Z}%
_2 $-grade 1, that is, they would very much look like a fermionic field.

Now, the $\bar \theta $'s generate their own Grassmann subalgebra of the
same dimension that the one generated by $\theta $'s; besides, we shall have
all the mixed products containing both types of generators, but which can be
always ordered e.g. with $\theta ^A$'s in front and $\bar \theta ^{\bar B}$%
's in the rear, by virtue of commutation relations. The products of $\theta
^A$'s alone or of $\bar \theta ^{\bar A}$'s alone span two subalgebras of
dimension $N(N+1)(N+2)/3$ each; the mixed products span new sectors of the $%
{\Bbb Z}_3$-graded Grassmann algebra.

{In the case of usual ${\Bbb Z}_2$-graded Grassmann algebras the
anti-commutation between the generators of the algebra and the assumed
associativity imply automatically the fact that {\it all} grade $0$ elements 
{\it commute} with the rest of the algebra, while {\it any two} elements of
grade $1$ anti-commute.}

{In the case of the ${\Bbb Z}_3$-graded generalization such an extension of
ternary and binary relations {\it does not follow automatically}, and must
be explicitly imposed. If we decide to extend the relations (\ref{z2grad0}),
(\ref{z2grad0bis}) and (\ref{z2grad1}) to {\it all} elements of the algebra
having a well-defined grade (i.e. the monomials in $\theta $'s and $\bar
\theta $'s), then many additional expressions must vanish, e.g.:}

\[
{\theta ^A\underbrace{{\theta ^B{\bar \theta }}^{\bar C}}=\underbrace{{%
\theta ^B{\bar \theta }}^{\bar C}}\theta ^A=\theta ^B\underbrace{{{\bar
\theta }^{\bar C}\theta }^{{A}}}={\bar \theta }^C\theta ^A\theta ^B=0} 
\]
{because on the one side, $\theta ^B{\bar \theta }^{\bar C}$ and ${\bar
\theta }^{\bar C}\theta ^A$ are of grade 0 and commute with all other
elements and on the other side, commuting ${\bar \theta }^C$ with $\theta
^A\theta ^B$ one gets twice the factor $j$, which leads to the overall
factor $j^2{\bar \theta }^C\theta ^A\theta ^B$. This produces a
contradiction which can be solved only by supposing that $\theta ^A\theta ^B{%
\bar \theta }^C=0$.}

{The resulting ${\Bbb Z}_3$-graded algebra contains only the following
combinations of generators: 
\[
A_1=\{\theta ,{\bar \theta }{\bar \theta }\}\text{; }A_2=\{{\bar \theta }%
,\theta \theta \}\text{; }A_0=\{{\bf 1},\theta {\bar \theta },\theta \theta
\theta ,{\bar \theta }{\bar \theta }{\bar \theta }\} 
\]
}

The dimension of the algebra is then 
\[
D(N)=1{}+2N+3N^2+\frac{2(N^3-N)}3=\frac{3+4N+9N^2+2N^3}3 
\]
The four summands $1$, $2N$, $3N^2$ and $\frac{2(N^3-N)}3$ correspond to the
subspaces respectively spanned by the combinations $\{{\Bbb C}\}$, $\{\theta
,\bar \theta \}$, $\{\theta \theta ,\theta \bar \theta ,\bar \theta \bar
\theta \}$ and $\{\theta \theta \theta ,\bar \theta \bar \theta \bar \theta
\}$.

{Let us note that the set of grade $0$ (which obviously forms a sub-algebra
of the ${\Bbb Z}_3$-graded Grassmann algebra) contains the products which
could symbolize the only observable combinations of {\it quark fields} in
quantum chromodynamics based on the $SU(3)$-symmetry.}

We can introduce the ${\Bbb Z}_3$-graded derivations of the ${\Bbb Z}_3$%
-graded Grassmann algebra by postulating the following set of rules\label
{grassderiv}: 
\[
\partial _A({\bf 1})=0\text{;\ \ }\partial _A\theta ^B={\delta }_A^B\text{;\
\ }\partial _A\bar \theta ^{\bar B}=0 
\]
and similarly 
\[
\partial _{\bar A}({\bf 1})=0\text{;\ }\ \partial _{\bar B}\bar \theta
^{\bar C}={\delta }_{\bar B}^{\bar C}\text{;\ \ }\partial _{\bar B}\theta
^A=0 
\]
When acting on various binary and ternary products, the derivation rules are
the following: 
\[
\partial _A(\theta ^B\theta ^C)={\delta }_A^B\theta ^C+j{\delta }_A^C\theta
^B\text{;\ \ }\partial _A(\theta ^B\theta ^C\theta ^D)={\delta }_A^B\theta
^C\theta ^D+j{\delta }_A^C\theta ^D\theta ^B+j^2{\delta }_A^D\theta ^B\theta
^C 
\]
Similarly for the conjugate entities: 
\[
\partial _{\bar A}(\bar \theta ^{\bar B}\bar \theta ^{\bar C})={\delta }%
_{\bar A}^{\bar B}\bar \theta ^{\bar C}+j^2{\delta }_{\bar A}^{\bar C}\bar
\theta ^{\bar B}\text{;\ \ }\partial _{\bar A}(\bar \theta ^{\bar B}\bar
\theta ^{\bar C}\bar \theta ^{\bar D})={\delta }_{\bar A}^{\bar B}\bar
\theta ^{\bar C}\bar \theta ^{\bar D}+j^2{\delta }_{\bar A}^{\bar C}\bar
\theta ^{\bar D}\bar \theta ^{\bar B}+j{\delta }_{\bar A}^{\bar D}\bar
\theta ^{\bar B}\bar \theta ^{\bar C} 
\]

Note the ``twisted'' Leibniz rule for the ternary products.

Finally, for mixed binary products like $\theta ^A\bar \theta ^{\bar B}$ the
derivation rules are as follows: 
\[
\partial _A(\theta ^B\bar \theta ^{\bar C})={\delta }_A^B\bar \theta ^{\bar
C}\text{; }{\partial }_{\bar A}(\theta ^B\bar \theta ^{\bar C})=j{\delta }%
_{\bar A}^{\bar C}{\theta }^B; 
\]
There is no need for rules of derivation of fourth-order homogeneous
expressions, because these vanish identically.

As the immediate consequence of these rules we have the following important
identities: 
\[
\partial _A\partial _B\partial _C=j\partial _B\partial _C\partial _A\text{
and }\partial _{\bar A}\partial _{\bar B}\partial _{\bar C}=j^2\partial
_{\bar B}\partial _{\bar C}\partial _{\bar A} 
\]
while 
\[
\partial _A\partial _{\bar C}=j\partial _{\bar C}\partial _A\text{ and }%
\partial _{\bar C}\partial _A=j^2\partial _A\partial _{\bar C} 
\]
hence the important consequence 
\begin{equation}
\partial _A\partial _B\partial _C+\partial _B\partial _C\partial _A+\partial
_C\partial _A\partial _B=0  \label{sum3deriv}
\end{equation}

\section{${\Bbb Z}_3$-graded algebra of hypersymmetry\protect\linebreak[4]
generators.}

The ${\Bbb Z}_3$-graded generalization of the Grassmanian and the ${\Bbb Z}%
_3 $-graded derivatives defined above can be used in order to produce a $%
{\Bbb Z}_3$-generalization of the supersymmetry generators acting on the
usual ${\Bbb Z}_2$-graded Grassmann algebra generated by anticommuting
fermionic variables ${\theta }^\alpha $ and $\bar \theta ^{\dot \beta }$ : 
\[
\theta ^\alpha \theta ^\beta +\theta ^\beta \theta ^\alpha =0\text{,\ \ }%
\bar \theta ^{\dot \alpha }\theta ^\beta +\theta ^\beta \bar \theta ^{\dot
\alpha }=0\text{,\ \ }\bar \theta ^{\dot \alpha }\bar \theta ^{\dot \beta
}+\bar \theta ^{\dot \beta }\bar \theta ^{\dot \alpha }=0 
\]
With the ``anti-Leibniz'' rule of derivation 
\[
\partial _\alpha (\theta ^\beta \theta ^\gamma )=\delta _\alpha ^\beta
\theta ^\gamma -\delta _\alpha ^\gamma \theta ^\beta 
\]
and similarly for any two dotted indices or mixed indices, one verifies
easily that all such derivations do anticommute: 
\[
\partial _\alpha \partial _\beta +\partial _\beta \partial _\alpha =0\text{%
,\ \ }\partial _{\dot \alpha }\partial _{\dot \beta }+\partial _{\dot \beta
}\partial _{\dot \alpha }=0\text{,\ \ }\partial _\alpha \partial _{\dot
\beta }+\partial _{\dot \beta }\partial _\alpha =0 \label{sum2deriv} 
\]

These rules enable us to construct the generators of the supersymmetric (or $%
{\Bbb Z}_2$-graded) ``odd'' translations:

\[
{\cal D}_\alpha =\partial _\alpha +{\sigma }_{\alpha \dot \beta }^k\bar
\theta ^{\dot \beta }\partial _k\text{,\ \ }{\cal D}_{\dot \beta }=\partial
_{\dot \beta }+{\sigma }_{\alpha \dot \beta }^m{\theta }^\alpha \partial _m 
\]
where both dotted and un-dotted indices $\alpha ,\dot \beta $ take the
values 1 and 2, while the space-time indices k,l,m run from 0 to 3. The
anti-commutators of these differential operators yield the ordinary
(``even'') space-time translations:

\[
{\cal D}_\alpha {\cal D}_{\dot \beta }+{\cal D}_{\dot \beta }{\cal D}_\alpha
=2\ \ {\sigma }_{\alpha \dot \beta }^k\partial _k 
\]

while 
\[
{\cal D}_\alpha {\cal D}_\beta +{\cal D}_\beta {\cal D}_\alpha =0\text{,\ \ }%
{\cal D}_{\dot \alpha }{\cal D}_{\dot \beta }+{\cal D}_{\dot \beta }{\cal D}%
_{\dot \alpha }=0 
\]

The ${\Bbb Z}_3$-graded generalization would amount to find a ``cubic root''
of a linear differential operator, making use of equation \ref{sum3deriv}.
We must have six kinds of generalized Grassmann variables $\theta ^A$, $%
\theta ^{\stackrel{\wedge }{A}}$, $\theta ^{\stackrel{\vee }{A}}$ on the one
hand and $\bar \theta ^{\bar A}$, $\bar \theta ^{\stackrel{\wedge }{\bar A}}$%
, $\bar \theta ^{\stackrel{\vee }{\bar A}}$ on the other hand, which is
formally analogous to the ${\Bbb Z}_2$-graded case. All kinds of $\theta $'s
and $\bar \theta $'s act like those that were introduced in section \ref
{z3grass}. Instead of the Pauli matrices we should introduce the entities
endowed with three indices (``cubic matrices'') with which the generators of
the ${\Bbb Z}_3$-graded translations of grade 1 and 2 may be constructed as
follows: 
\[
{\it D}_A=\partial _A+{\rho }_{A\stackrel{\wedge }{B}\stackrel{\vee }{C}%
}^m\theta ^{\stackrel{\wedge }{B}}\theta ^{\stackrel{\vee }{C}}{\nabla }%
_m+\omega _{A\bar A}^m\bar \theta ^{\bar A}{\nabla }_m\text{,\ \ }{\it D}%
_{\bar A}=\partial _{\bar A}+{\bar \rho }_{\bar A\stackrel{\wedge }{\bar B}%
\stackrel{\vee }{\bar C}}^m\bar \theta ^{\stackrel{\wedge }{\bar B}}\bar
\theta ^{\stackrel{\vee }{\bar C}}{\nabla }_m+\bar \omega _{\bar AA}^m\theta
^A{\nabla }_m 
\]
\[
{\it D}_{\stackrel{\wedge }{B}}=\partial _{\stackrel{\wedge }{B}}+{\rho }_{A%
\stackrel{\wedge }{B}\stackrel{\vee }{C}}^m\theta ^A\theta ^{\stackrel{\vee 
}{C}}{\nabla }_m+\omega _{\stackrel{\wedge }{B}\stackrel{\wedge }{\bar B}%
}^m\bar \theta ^{\stackrel{\wedge }{\bar B}}{\nabla }_m\text{,\ \ }{\it D}_{%
\stackrel{\wedge }{\bar B}}=\partial _{\stackrel{\wedge }{\bar B}}+{\bar
\rho }_{\bar A\stackrel{\wedge }{\bar B}\stackrel{\vee }{\bar C}}^m\bar
\theta ^{\bar A}\bar \theta ^{\stackrel{\vee }{\bar C}}{\nabla }_m+\bar
\omega _{\stackrel{\wedge }{\bar B}\stackrel{\wedge }{B}}^m\theta ^{%
\stackrel{\wedge }{B}}{\nabla }_m 
\]
\[
{\it D}_{\stackrel{\vee }{C}}=\partial _{\stackrel{\vee }{C}}+{\rho }_{A%
\stackrel{\wedge }{B}\stackrel{\vee }{C}}^m\theta ^A\theta ^{\stackrel{%
\wedge }{B}}{\nabla }_m+\omega _{\stackrel{\vee }{C}\stackrel{\vee }{\bar C}%
}^m\bar \theta ^{\stackrel{\vee }{\bar C}}{\nabla }_m\text{,\ \ }{\it D}_{%
\stackrel{\vee }{\bar C}}=\partial _{\stackrel{\vee }{\bar C}}+{\bar \rho }%
_{\bar A\stackrel{\wedge }{\bar B}\stackrel{\vee }{\bar C}}^m\bar \theta
^{\bar A}\bar \theta ^{\stackrel{\wedge }{\bar B}}{\nabla }_m+\bar \omega _{%
\stackrel{\vee }{\bar C}\stackrel{\vee }{C}}^m\theta ^{\stackrel{\vee }{C}}{%
\nabla }_m 
\]

The nature of the indices needs not to be specified; the only important
thing to be assumed at this stage is that the differential operators $\nabla
_m$ do commute with the ${\Bbb Z}_3$-graded differentiations $\partial _A$.
It is also interesting to consider the operators one gets when the $\nabla
_m $ are replaced with {\em supersymmetric} derivations (that anticommute
with the ${\Bbb Z}_3$-graded differentiations). But in the simpler case
described here, the following operators acting on the ${\Bbb Z}_3$-graded
generalized Grassmanian: 
\begin{eqnarray*}
D_{ABC}^{III} &=&{\it D}_A{\it D}_B{\it D}_C+{\it D}_B{\it D}_C{\it D}_A+%
{\it D}_C{\it D}_A{\it D}_B+{\it D}_C{\it D}_B{\it D}_A+{\it D}_B{\it D}_A%
{\it D}_C+{\it D}_A{\it D}_C{\it D}_B \\
\bar D_{\bar A\bar B\bar C}^{III} &=&{\it D}_{\bar A}{\it D}_{\bar B}{\it D}%
_{\bar C}+{\it D}_{\bar B}{\it D}_{\bar C}{\it D}_{\bar A}+{\it D}_{\bar C}%
{\it D}_{\bar A}{\it D}_{\bar B}+{\it D}_{\bar C}{\it D}_{\bar B}{\it D}%
_{\bar A}+{\it D}_{\bar B}{\it D}_{\bar A}{\it D}_{\bar C}+{\it D}_{\bar A}%
{\it D}_{\bar C}{\it D}_{\bar B} \\
D_{A\bar A}^{II} &=&{\it D}_A{\it D}_{\bar A}-j^2{\it D}_{\bar A}{\it D}_A
\end{eqnarray*}
represent {\it homogeneous} operators on the ${\Bbb Z}_3$-graded Grassmann
algebra, i.e. they map polynomials in $\theta $'s of a given grade into
polynomials of the same grade; the result can be represented by a
complex-valued matrix containing various combinations of the
differentiations $\nabla _m$ ; their eventual symmetry properties will
depend on the assumed symmetry properties of the matrices $\rho _{ABC}$ and $%
\omega _{A\bar B}$.

Let us consider in more detail the case of dimension $3$ (the simplest
possible realization of the ${\Bbb Z}_3$-graded Grassmannian and the
derivations on it is of course the case with {\it one} generator and its
conjugate; this has been considered in a paper by Won-Sang Chung\cite
{wonsangchung}).

The dimension of the ${\Bbb Z}_3$-graded Grassmann algebra with three grade-$%
1$ generators $\theta $, $\stackrel{\wedge }{\theta }$ and $\stackrel{\vee }{%
\theta }$ and three ``conjugate'' grade-$2$ generators $\bar \theta $, $%
\stackrel{\wedge }{\bar \theta }$ and $\stackrel{\vee }{\bar \theta }$ is $%
51 $; any linear operator, including the derivations $\partial _A$ and the
multiplication by any combination of the generators, as well as the
operators ${\it D}_A$ and ${\it D}_{\bar A}$ introduced above, can be
represented by means of $51\times 51$ complex-valued matrices.
Unfortunately, the operators $D^{II}$ and $D^{III}$ are neither diagonal nor
diagonalizable. But if we apply them to a scalar function $f$, we get: 
\[
D_{1\bar 1}^{II}f=(\omega _{1\bar 1}^m+\bar \omega _{\bar 11}^m)\nabla _mf 
\]
and 
\[
D_{1\stackrel{\wedge }{1}\stackrel{\vee }{1}}^{III}f=-3j^2\rho _{1\stackrel{%
\wedge }{1}\stackrel{\vee }{1}}^m\nabla _mf\text{ as well as }\bar D_{\bar 1%
\stackrel{\wedge }{\bar 1}\stackrel{\vee }{\bar 1}}^{III}f=-3j\bar \rho
_{\bar 1\stackrel{\wedge }{\bar 1}\stackrel{\vee }{\bar 1}}^m\nabla _mf 
\]
The $\omega $ matrices are the only ones that remain in the $D^{II}$ whereas
the $\rho $ cubic matrices emerge from the ternary combinations $D^{III}$.
On the space of scalar functions, our operators act simultaneously{\em \ }as 
{\em square }and {\em cubic} roots of ordinary translations. Using
extensions of these objects where the ${\nabla }_m$ are replaced with the
supersymmetry generators, we have contructed a simple ${\Bbb Z}_3$-graded
non commutative geometry model featuring three Higgs fields. The lagrangian
contains the potential term of degree 6: 
\[
V=3\left| \Phi _1+\Phi _2+\Phi _3+\Phi _1\Phi _2+\Phi _2\Phi _3+\Phi _3\Phi
_1+\Phi _1\Phi _2\Phi _3\right| ^2 
\]
and implies multiple spontaneous symmetry breaking. This model will be the
subject of another article.

\section{${\Bbb Z}_3$-graded matrices.}

If we want to have an integration theory on ``hypermanifolds'', we will need
the equivalents of matrices and determinant that should naturally appear in
the formula of change of variables. If we align the basis of our ${\Bbb Z}_3$%
-graded Grassmann algebra, with all the elements of grade $0$ first, then
all the elements of grade $1$ and finally the elements of grade $2$ in a
one-column vector, any linear transformation that would leave these entries
in the same order can be symbolized like the $D^{II}$ and $D^{III}$
operators by a matrix whose entries have a definite ${\Bbb Z}_3$-grade
placed as follows: 
\[
\left( 
\begin{array}{ccc}
0 & 2 & 1 \\ 
1 & 0 & 2 \\ 
2 & 1 & 0
\end{array}
\right) \left( 
\begin{array}{c}
0 \\ 
1 \\ 
2
\end{array}
\right) =\left( 
\begin{array}{c}
0 \\ 
1 \\ 
2
\end{array}
\right) 
\]
then the position of the three grades does not change in the resulting
column; we shall call any matrix displaying this block grading structure a 
{\it grade 0} matrix. We can introduce two other kinds of matrices that
raise all the grades by $1$ resp. by $2$, like the $D_A$ resp. $D_{\bar A}$
operators, as follows and calling them respectively {\it grade 1} and {\it %
grade 2} matrices: 
\[
\left( 
\begin{array}{ccc}
1 & 0 & 2 \\ 
2 & 1 & 0 \\ 
0 & 2 & 1
\end{array}
\right) \left( 
\begin{array}{c}
0 \\ 
1 \\ 
2
\end{array}
\right) =\left( 
\begin{array}{c}
1 \\ 
2 \\ 
0
\end{array}
\right) \text{ and }\left( 
\begin{array}{ccc}
2 & 1 & 0 \\ 
0 & 2 & 1 \\ 
1 & 0 & 2
\end{array}
\right) \left( 
\begin{array}{c}
0 \\ 
1 \\ 
2
\end{array}
\right) =\left( 
\begin{array}{c}
2 \\ 
0 \\ 
1
\end{array}
\right) 
\]
(the numbers $0,1,2$ symbolizing the grades of the respective entries in the
matrices). These matrices have been studied in detail in \cite{z3mat}.

It is easy to check that the grades of these matrices add up modulo $3$
under the associative matrix multiplication law.

One can define the analogues of the {\em supertrace} and {\em %
superdeterminant} of a matrix 
\[
M=\left( 
\begin{array}{ccc}
A & B & C \\ 
D & E & F \\ 
G & H & I
\end{array}
\right) 
\]
as follows:\label{z3trace} 
\[
\limfunc{tr}\nolimits_{{\Bbb Z}_3}(M)=\limfunc{tr}A+j^{2(1-gr(M))}\limfunc{tr%
}E+j^{(1-gr(M))}\limfunc{tr}I 
\]
and, for grade $0$ matrices:

\[
\begin{array}{r}
\det\nolimits_{{\Bbb Z}_3}(M)=\det
(A-CI^{-1}G-(B-CI^{-1}H)(E-FI^{-1}H)^{-1}(D-FI^{-1}G))\times \\ 
\times (\det (E-FI^{-1}H))^{j^2}(\det I)^j
\end{array}
\]
There are five other equivalent formulations of the ${\Bbb Z}_3$-determinant
reflecting the six-fold $S_3$ symmetry, just as there are two alternative
formulations of the superdeterminant that reflect the $S_2$ symmetry\cite
{kapranov}.

The ${\Bbb Z}_3$ trace and determinant satisfy all the usual properties one
would expect, especially the most important one: 
\[
\det\nolimits_{{\Bbb Z}_3}(\exp (M))=\exp (\limfunc{tr}\nolimits_{{\Bbb Z}%
_3}(M)) 
\]
We expect this determinant to play the same role in integration theory on
``hypermanifolds'' as the superdeterminant in integration on supermanifolds.

\section{The ${\Bbb Z}_3$-graded exterior differential.}

Consider the algebra $M_3({\Bbb C})$ of $3\times 3$ complex matrices, with
the ${\Bbb Z}_3$-grading introduced above. Let $B,C$ denote two matrices
whose grades are $grad(A)=a$ and $grad(B)=b$, respectively. We define the 
{\it ${\Bbb Z}_3$-graded commutator} $[B,C]$ as follows: 
\[
\lbrack B,C]_{{\Bbb Z}_3}:=BC-j^{bc}CB, 
\]
(Note that this ${\Bbb Z}_3$-graded commutator does not satisfy the Jacobi
identity). Let $\eta $ be a matrix of grade $1$; we can choose for the sake
of simplicity 
\[
\eta =\left( 
\begin{array}{ccc}
0 & 1 & 0 \\ 
0 & 0 & 1 \\ 
1 & 0 & 0
\end{array}
\right) 
\]

With the help of the matrix $\eta $ we can define a formal ``differential''
on the ${\Bbb Z}_3$-graded algebra of $3\times 3$ matrices as follows: 
\[
dB:=[\eta ,B]_{{\Bbb Z}_3}=\eta B-j^bB\eta 
\]

It is easy to show that $d(BC)=(dB)C+j^bB(dC)$ and that $d^3=0$. It is also
possible to define an external differential with a similar property, {\it %
i.e.} $d^2\neq 0$, $d^3=0$, on a manifold.

Let $M_n$ be a differentiable manifold of dimension $N$, with local
coordinates ${x^k}$. The variables $x^k$ commute and their ${\Bbb Z}_3$%
-grade is $0$. The linear operator $d$ applied to $x^k$ produces a 1-form
whose ${\Bbb Z}_3$-grade is $1$ by definition; when applied two times by
iteration, it produces a new form of grade $2$, denoted by $d^2x^k$. We
shall postulate $d^3=0$.

Let $F(M)$ denote the algebra of functions $C^\infty (M)$, over which the $%
{\Bbb Z}_3$-graded algebra generated by the forms $dx^i$ and $d^2x^k$
behaves as a {\it left} module. In other words, we shall be able to multiply
the forms $dx^i$ , $d^2x^k$, $dx^idx^k$ , etc. by the functions {\it only on
the left}; right multiplication will just not be considered here. That is
why we will write by definition, e.g. 
\[
d(x^ix^k):=x^idx^k+x^kdx^i 
\]

We shall also assume the following Leibniz rule for the operator $d$ with
respect to the multiplication of the ${\Bbb Z}_3$-graded forms: when $d$
crosses a form of grade $p$, the factor $j^p$ appears as follows: 
\[
d(\omega \ \ \phi )=(d\omega )\phi +j^p\omega (d\phi ) 
\]

Let us note that in contrast with the ${\Bbb Z}_2$-graded case, the forms
are treated as one whole, even when multiplied from the left by an arbitrary
function; that means that we can not identify e.g. $(\omega _idx^i)(\phi
_kdx^k)$ with $(\omega _i\phi _k)dx^idx^k$

This is equivalent with saying that the products of functions by the forms
are to be understood in the sense of tensor products, which is associative,
but non-commutative.

Nevertheless, such an identification can be done for the forms of maximal
degree (i.e. $3$), which contain the products of the type $dx^idx^kdx^m$ or $%
dx^id^2x^m$, whose exterior differentials vanish disrespectfully of the
order of the multiplication.

With the so established ${\Bbb Z}_3$-graded Leibniz rule, the postulate $%
d^3=0$ suggests in an almost unique way the ternary and binary commutation
rules for the differentials $dx^i$ et $d^2x^k$. To begin with, consider the
differentials of a function of the coordinates $x^k$, with the ``first
differential'' $df$ that coincides with the usual one: 
\begin{eqnarray*}
df &:&=(\partial _if)dx^i \\
d^2f &:&=(\partial _k\partial _if)dx^kdx^i+(\partial _if)d^2x^i \\
d^3f &=&(\partial _m\partial _k\partial _if)dx^mdx^kdx^i+(\partial
_k\partial _if)d^2x^kdx^i+j(\partial _k\partial _if)dx^id^2x^k+(\partial
_k\partial _if)dx^kd^2x^i
\end{eqnarray*}
(we remind that the last part of the differential, $(\partial _if)d^3x^i$,
vanishes by virtue of the postulate $d^3x^i=0$).

Supposing that the partial derivatives commute, exchanging the summation
indices $i$ et $k$ in the last expression and replacing $1+j$ by $-j^2$, we
arrive at the following two independant conditions that lead to the
vanishing of $d^3f$ : 
\[
dx^mdx^kdx^i+dx^kdx^idx^m+dx^idx^mdx^k=0\text{\qquad and\qquad }%
d^2x^kdx^i-j^2dx^id^2x^k=0\ \ . 
\]
which lead in turn to the following choice of relations (which of course is
not unique): 
\[
dx^idx^kdx^m=jdx^kdx^mdx^i\text{\qquad and\qquad }dx^id^2x^k=jd^2x^kdx^i. 
\]

By extending these rules to {\it all} the expressions with a well-defined
grade, and applying the associativity of the ${\Bbb Z}_3$-exterior product,
it is easy to verify that all the expressions of the type $dx^idx^kdx^mdx^n$
and $dx^idx^kd^2x^m$ must vanish, and along with these, also the monomials
of higher order that would contain them as factors. Still, this is not
sufficient in order to satisfy the rule $d^3=0$ on all the forms spanned by
the generators $dx^k$ and $d^2x^k$. It can be proved without much effort
that the expressions containing $d^2x^id^2x^k$ must vanish, too. For
example, if we take the particular 1-form $x^idx^k$ and apply to it the
operator $d$, we get 
\[
d(x^idx^k)=dx^idx^k+x^id^2x^k; 
\]
\[
d^2(x^idx^k)=d^2x^idx^k+(1+j)dx^id^2x^k=d^2x^idx^k-d^2x^kdx^i; 
\]
which leads to $d^3(x^idx^k)=d^2x^id^2x^k-d^2x^kd^2x^i$. There is another
possibility which is to say that the $d^2x^k$ should commute with one another%
\footnote{%
this has been suggested by C. Juszczak}. This makes the differential algebra
infinite, but prevents us from considering the $d^2x^k$ as grade $2$
entities. But if we want to keep both the associativity of the ``exterior
product'' and the ternary rule for the entities of grade $2$, i.e. $%
d^2x^id^2x^kd^2x^m=j^2d^2x^kd^2x^md^2x^i$, then the only solution is to
impose $d^2x^id^2x^k=0$ and to set forward the additional rule declaring
that any expression containing {\it fourth or higher} power of the operator $%
d$ must vanish identically.

With the above set of rules we can check that $d^3=0$ on all the forms,
whatever their grade or degree. Let us show how such calculus works on the
example of a 1-form $\omega =\omega _kdx^k$: 
\[
d(\omega _kdx^k)=(\partial _i\omega _k)dx^idx^k+\omega _kd^2x^k; 
\]
\[
d^2(\omega _kdx^k)=(\partial _m\partial _i\omega _k)dx^mdx^idx^k+(\partial
_i\omega _k)d^2x^idx^k+jdx^id^2x^k+\partial _i\omega _kdx^id^2x^k; 
\]
after exchanging the summation indices $i$ and $k$ in two last terms and
using the fact that $j+1=-j^2$ and the commutation relations between $dx^k$
and $d^2x^i$, we can write 
\[
d^2(\omega _kdx^k)=(\partial _m\partial _i\omega _k)dx^mdx^idx^k+(\partial
_i\omega _k-\partial _k\omega _i)d^2x^idx^k. 
\]
where it is interesting to note how the usual {\em anti-symmetric} exterior
differential appears as a part of the whole expression.

It is also easy to check that $Im(d)\subseteq Ker(d^2)$ and $%
Im(d^2)\subseteq Ker(d)$.

\section{${\Bbb Z}_3$-graded gauge theory.}

Let {\cal A} be an associative algebra with unit element, and let {\cal H}
be a free left module over this algebra. Let $A$ be a {\cal A}-valued 1-form
defined on a differential manifold $M$, and let $\Phi$ be a function on the
manifold $M$ with values in the module {\cal H}.

We shall introduce the {\it covariant differential} as usual: 
\[
D\Phi :=d\Phi +A\Phi ; 
\]

If the module is a free one, any of its elements $\Phi$ can be represented
by an appropriate element of the algebra acting on a fixed element of {\cal H%
}, so that one can always write $\Phi = B \Phi_o $; then the action of the
group of automorphisms of {\cal H} can be translated as the action of the
same group on the algebra {\cal A}.

Let $U$ be a function defined on $M$ with its values in the group of the
automorphisms of {\cal H}. The definition of a covariant differential is
equivalent with the requirement $DU^{-1}B=U^{-1}DB$; as in the usual case,
this leads to the following well-known transformation for the connection
1-form $A$ : 
\[
A\Rightarrow U^{-1}AU+U^{-1}dU; 
\]

But here, unlike in the usual theory, the second covariant differential $%
D^2\Phi $ is not an automorphism: as a matter of fact, we have: 
\[
D^2\Phi =d(d\Phi +A\Phi )+A(d\Phi +A\Phi )=d^2\Phi +dA\Phi +jAd\Phi +Ad\Phi
+A^2\Phi ; 
\]
the expression containing $d^2\Phi $ et $d\Phi $ ; whereas $D^3\Phi $ is an
automorphism indeed, because it contains only $\Phi $ multiplied on the left
by an algebra-valued 3-form: 
\[
D^3\Phi =(d^2A+d(A^2)+AdA+A^3)\Phi =(D^2A)\Phi :=\Omega \Phi ; 
\]

Obviously, because $D (U^{-1} \Phi) = U^{-1} (D \Phi) $, one also has:

$D^3(U^{-1}\Phi )=U^{-1}(D^3\Phi )=U^{-1}\Omega \Phi =U^{-1}\Omega
UU^{-1}\Phi $, which proves that the 3-form $\Omega $ transforms as usual, $%
\Omega \longmapsto U^{-1}\Omega U$ when the connection 1-form transforms
according to the law: $A\longmapsto U^{-1}AU+U^{-1}dU$.

It can be also proved by a direct calculus that the curvature 3-form $\Omega 
$ does vanish identically for $A=U^{-1}dU$ (see \cite{z3diff}).

It is interesting to compute the expression of the curvature 3-form in local
coordinates: 
\[
\Omega =d^2A+d(A^2)+AdA+A^3=\Omega _{ikm}dx^idx^kdx^m+F_{ik}d^2x^idx^k 
\]
where 
\[
\Omega _{ikm}:=\partial _i\partial _kA_m+A_i\partial _kA_m-\partial
_kA_mA_i+A_iA_kA_m\text{\quad and\quad }F_{ik}:=\partial _iA_k-\partial
_kA_i+A_iA_k-A_kA_i 
\]
In $F_{ik}$ one can easily recognize the 2-form of curvature of the {\em %
usual} gauge theories.

We know that the expression $F_{ik}$ is covariant with respect to the gauge
transformations; on the other hand, the 3-form $\Omega $ is also covariant;
therefore, the local expression $\Omega _{ijk}$ must be covariant, too. As a
matter of fact, it can be expressed as a combination of covariant
derivatives of the 2-form $F_{ik}$: 
\[
\Omega _{ikm}=\frac 13[jD_iF_{mk}+j^2D_kF_{mi}], 
\]
or, equivalently, 
\[
\Omega _{ikm}=-\frac 16[D_iF_{mk}+D_kF_{mi}]+\frac{i\sqrt{3}}%
6[D_iF_{mk}-D_kF_{mi}] 
\]

The natural symmetry between $j$ et $j^2$ , which leads to the possibility
of choosing one of these two complex numbers as the generator of the group $%
{\Bbb Z}_3$ , and simultaneous interchanging the roles between the grades $1$
and $2$, suggests that we could extend the notion of complex conjugation $%
j\Rightarrow (j)^{*}:=j^2$, with $((j)^{*})^{*}=j$, to the algebra of ${\Bbb %
Z}_3$- graded exterior forms and the operator $d$ itself.

It does not seem reasonable to use the ``second differentials'' $d^2x^i$ as
the objects conjugate to the ``first differentials'' $dx^i$, because the
rules of ${\Bbb Z}_3$-graded exterior differentiation we have imposed break
the symmetry between these two kinds of differentials: remember that the
products $dx^idx^k$, and $dx^idx^kdx^m$ are admitted, while we require that $%
d^2x^id^2x^k$ and $d^2x^id^2x^kd^2x^m$ must vanish.

This suggests the introduction of a ``{\it conjugate}'' differential $\delta 
$ of grade $2$, the image of the differential $d$ under the conjugation $*$,
satisfying the following conjugate relations: 
\[
\delta x^i\delta x^k\delta x^m=j^2\delta x^k\delta x^m\delta x^i\text{, }%
\delta x^i\delta ^2x^k=j^2\delta ^2x^k\delta x^i\ \ . 
\]

All the relations existing between the operator $d$ and the exterior forms
generated by $dx^i$ and $d^2x^k$ are faithfully reproduced under the
conjugation $*$ if we consider the ${\Bbb Z}_3$-graded algebra generated by
the entities $\delta x^i$ and $\delta ^2x^k$ as a {\it right module} over
the algebra of functions $F(M)$, with the operator $\delta $ acting {\it on
the right} on this module.

The rules $d^3=0$ and $\delta ^3=0$ suggest their natural extension: 
\[
d\delta =\delta d=0 
\]

We would like to be able to form quadratic expressions that could define a
scalar product; to do this, we should postulate that the algebra generated
by the elements $dx^i$, $d^2x^k$ and its conjugate algebra generated by the
elements $\delta x^i$, $\delta x^k$ commute with each other.

Then, we can define scalar products for the forms of maximal degree $3$:%
\linebreak[4] $<\omega \mid \phi >:=*\omega \phi $ , and integrating this
result with respect to the usual volume element defined on the manifold $M$,
which gives explicitly: 
\[
\int {\bar \omega _{ikm}}\phi _{prs}<\delta x^i\delta x^k\delta x^m\mid
dx^pdx^rdx^s>\quad \text{and\quad }\int {\bar \psi _{ik}}\chi _{mn}<\delta
x^i\delta ^2x^k\mid d^2x^mdx^n> 
\]

What remains now is to determine the scalar products of the basis of forms;
in order to assure the hermiticity of the product, one can always choose an
``orthonormal'' basis in which we should have: 
\[
<\delta x^i\delta x^k\delta x^m\mid dx^pdx^rdx^s>=\delta _s^i\delta
_r^k\delta _p^m\text{\quad and\quad }<\delta x^i\delta ^2x^k\mid
d^2x^mdx^n>=\mu \delta _n^i\delta _m^k. 
\]

Here the first scalar product is normed to $1$, and $\mu $ is the ratio
between the two types of ``elementary volume''.

We consider the two types of forms of degree $3$, $dx^idx^kdx^m$ and $%
d^2x^pdx^r$, as being mutually orthogonal.

The above scalar product and generalized integral of ${\Bbb Z}_3$-graded
forms enable us to introduce the Lagrangian densities involving quadratic
expressions in both $F_{ik}$ and $\Omega _{ikl}$.

\section{A ternary generalization of Lie algebras.}

Once the ``ternary logic'' is adopted, it can be quite easily applied to
other well-known algebraic concepts, e.g. Lie algebras. There are other
ternary generalizations of Lie algebra (see for example \cite{nambu,
takhtajan, trilie, trilie2}), from which this one is quite different, as far
as we know.

The well-known Ado's theorem states that any finite-dimensional Lie algebra
can be realized with the elements of an associative algebra (the so-called
enveloping algebra, whose dimension is usually much bigger that the
dimension of the Lie algebra we started with) so that the skew-symmetric
(and non-associative) multiplication law of the Lie algebra is replaced by
the commutator $[A,B]=AB-BA$. The Jacobi identity, which is an independent
postulate in the definition of an abstract Lie algebra, follows then
automatically due to the associativity in the enveloping algebra.

The skew-symmetric composition law defined by a commutator in an associative
algebra can be readily and naturally generalized as follows: 
\[
\{X,Y,Z\}=XYZ+jYZX+\ j^2ZXY+ZYX+j^2YXZ+jXZY 
\]

The ternary ``3-commutator'' thus defined satisfies the following properties
corresponding to the antisymmetry property in the usual case: 
\[
\{X,Y,Z\}=j\{Y,Z,X\}=j^2\{Z,X,Y\} 
\]
from which follow the {\it two} identities: 
\[
\{X,Y,Z\}+\{Y,Z,X\}+\{Z,X,Y\}=0\text{\quad and\quad }\{X,Y,Z\}+j\{Y,Z,X%
\}+j^2\{Z,X,Y\}=0 
\]

A very simple result seems very significant here: any finite ternary
algebraic structure defined above contains and induces the {\it ordinary Lie
algebra structure} if the associative algebra includes the unit element
(denoted by $1$ here) commuting with all other elements. Here is the proof: 
\begin{eqnarray*}
\{X,1,Y\} &=&(X1Y)+j(1YX)+j^2(YX1)= \\
&=&(XY)+j(YX)+j^2(YX)=XY-YX=[X,Y]
\end{eqnarray*}

because $j + j^2 = - 1$

Here again, the Jacobi identity can be reconstructed if we iterate the same
ternary bracket with the unit element employed twice: 
\[
\{\{X,1,Y\},1,Z\}+\{\{Y,1,Z\},1,X\}+\{\{Z,1,X\},1,Y\}= 
\]
\[
=\ [[X,Y],Z]+[[Y,Z],X]+[[Z,X],Y]=0 
\]
Nevertheless the straightforward ternary generalization of the Jacobi
identity is not satisfied if we replace the unity $1$ by arbitrary elements $%
S$ and $T$ of the associative algebra: 
\[
\{\{X,S,Y\},T,Z\}+\{\{Y,S,Z\},T,X\}+\{\{Z,S,X\},T,Y\}\not =0 
\]

The Jacobi identity generalizes in fact in two distinct identities: 
\[
\{Z,\{X,S,T\},Y\}+\{X,\{S,Y,T\},Z\}+\{Y,\{S,T,Z\},X\}+\{S,\{X,Y,Z\},T\}= 
\]
\[
=\{Z,\{X,T,S\},Y\}+\{X,\{T,Y,S\},Z\}+\{Y,\{T,S,Z\},X\}+\{T,\{X,Y,Z\},S\} 
\]
and 
\[
\{T,\{X,Y,Z\},S\}+\{T,\{Y,X,S\},Z\}+\{T,\{Z,S,X\},Y\}+\{T,\{S,Z,Y\},X\}= 
\]
\[
=\{S,\{Z,X,Y\},T\}+\{Z,\{S,Y,X\},T\}+\{Y,\{X,Z,S\},T\}+\{X,\{Y,S,Z\},T\} 
\]

We can explore the structure of the simplest ternary algebras defined by the
abstract ternary bracket displaying the well-defined symmetry property $%
\{X,Y,Z\}=j\{Y,Z,X\}$. We shall start with an example showing how such a
non-associative ternary composition rule can be introduced even without any
associative algebra behind.

Let us consider the linear space of the {\it tri-linear forms} over a given
N-dimensional vector space E over complex numbers. In a given basis, any
such 3-form can be represented by its components $U_{ABC}$, $A,B..=1,...,N$.
Suppose also that a metric is defined on E; we can always diagonalize it and
choose the basis in which its components form the Kronecker tensor $\delta
_{AB}$; its inverse is then $\delta ^{BC}$ . Now, with the help of the
metric we can raise the indices, and the following ternary composition rule
for the 3-forms may be defined: 
\[
(U,V,W)_{ABC}=\delta ^{EF}U_{FAG}\delta ^{GH}V_{HBJ}\delta ^{JK}W_{KCE}\ \ ; 
\]
In what follows, we shall just sum over the repeated indices without writing
explicitly the metric: 
\[
(U,V,W)_{ABC}=U_{EAG}V_{GBH}W_{HCE} 
\]
This is the closest analogue of matrix multiplication that we can imagine,
although it is non-associative, what can be readily verified by definition.
Now, in order to come closer to ternary analogue of a Lie algebra, we should
form a ``ternary commutator'' as introduced above: 
\[
\{U,V,W\}_{ABC}=(U,V,W)_{ABC}+j(V,W,U)_{ABC}+j^2(W,U,V)_{ABC} 
\]
Obviously, the resulting 3-form displays the following internal symmetry: 
\[
\{U,V,W\}_{ABC}=j\{U,V,W\}_{BCA}=j^2\{U,V,W\}_{CAB} 
\]
If we want our ternary ${\Bbb Z}_3$-graded algebra to be closed under this
new composition law, we should restraint it to the ternary forms having the
above symmetry property. Such 3-forms will span a vector space over {\it %
real numbers}; its (real) dimension is $(N^3-N)/3$.

The most primitive ternary algebra of such 3-forms is obtained when the
indices $A,B,\ldots $ take on two values only, $1$ and $2$. The only two
independent elements of this algebra are the following: 
\[
\rho _{121}^{(1)}=1\text{,\qquad }\rho _{211}^{(1)}=j^2\text{,\qquad }\rho
_{112}^{(1)}=j 
\]
all other components vanishing, and 
\[
\rho _{212}^{(2)}=1\text{,\qquad }\rho _{122}^{(2)}=j^2\text{,\qquad }\rho
_{221}^{(2)}=j 
\]
with all other components vanishing.

Then the following ternary algebra is generated by these two 3-forms: 
\[
\{\rho ^{(1)},\rho ^{(2)},\rho ^{(1)}\}=-\rho ^{(2)}\text{\quad and\quad }%
\{\rho ^{(2)},\rho ^{(1)},\rho ^{(2)}\}=-\rho ^{(1)}. 
\]

By definition, the ternary commutator of the same 3-form taken with itself
three times is automatically vanishing.

Suppose now that we wish to represent this algebra by means of associative
complex matrices, with the ternary composition law defined as 
\[
\{A,B,C\}=ABC+jBCA+j^2CAB+CBA+j^2BAC+jACB 
\]

In the case of the ternary algebra of the 3-forms $\rho ^\alpha $ defined
above, with\linebreak[4] $\alpha =1,2$, complex $2\times 2$-matrices are
enough to define the only two elements. It is a simple exercise to prove
that such a realization is given by any two Pauli matrices (e.g. $\sigma
_3,\sigma _2$) divided by $\sqrt{2}$.

It is also possible to represent the eight dimensional ternary algebra
obtained when the indices $A,B,\ldots $ take three values by means of $%
3\times 3$ complex matrices. These may be viewed as the generators of the
Lie algebra of the unitary group $SU(3)$ , although in a somewhat unusual
basis.

\section{Ternary generalization of Clifford algebras.}

Let us rewrite the usual definition of Clifford algebra 
\[
\gamma ^\mu \gamma ^\nu +\gamma ^\nu \gamma ^\mu =2g^{\mu \nu }\ \ 1 
\]
with $\mu ,\nu =1,2,\ldots ,N$, in a slightly different way, which will
immediately suggest the ${\Bbb Z}_3$-graded generalization: 
\[
\gamma ^\mu \gamma ^\nu =-\gamma ^\nu \gamma ^\mu +2g^{\mu \nu }\ \ 1 
\]

Let us consider an algebra spanned by $N$ generators $Q^k$ whose ternary
products should satisfy the identity

\[
Q^{k_1}Q^{k_2}Q^{k_3}=\varrho _{k_1k_2k_3}(\sigma )\,Q^{k_{\sigma
(1)}}Q^{k_{\sigma (2)}}Q^{k_{\sigma (3)}}+3\,{\eta }^{k_1k_2k_3}\,1, 
\]
where $k_i=1,\ldots ,N$, $\sigma $ is the substitution $(1\,2\,3)\in {\Bbb Z}%
_3$ and $\varrho _{k_1k_2k_3}$ is the representation of the cyclic group $%
{\Bbb Z}_3$ by complex numbers which depend on the indices $k_1,k_2,k_3$ ($%
\varrho _{k_1k_2k_3}(\sigma )=j$ or $\varrho _{k_1k_2k_3}(\sigma )=j^2$) and
do not change under the cyclic permutation of the indices, i.e. $\varrho
_{k_1k_2k_3}=\varrho _{k_2k_3k_1}=\varrho _{k_3k_1k_2}$. If all binary
products $Q^kQ^m$ are linearly independent we shall call the algebra
generated by $Q^1,Q^2,\ldots ,Q^N$ {\em ternary} Clifford algebra. In the
case where the above identity for ternary products can be derived from some
binary identities, we shall call the corresponding algebra ${\Bbb Z}_3$%
-graded Clifford algebra.

Applying the above relation three times in a row one obtains the condition
on the tensor $\eta^{klm}$:

\[
{\eta }^{klm}+\varrho _{klm}(\sigma ){\eta }^{lmk}+\varrho _{klm}^2(\sigma ){%
\eta }^{mkl}=0. 
\]
Since $1+\varrho _{klm}+\varrho _{klm}^2=0$ the above equation has two
independent solutions, namely

\[
\text{(a):\ }\eta ^{klm}=\eta ^{lmk}=\eta ^{mkl}\text{ and (b):\ }\eta
^{klm}=\varrho _{klm}^2(\sigma )\eta ^{lmk}=\varrho _{klm}(\sigma )\eta
^{mkl}. 
\]
The first choice (a) implies the following relation for the generators $Q^m$:

\[
Q^kQ^lQ^m+Q^lQ^mQ^k+Q^mQ^kQ^l=3(1-\varrho _{klm}^2(\sigma ))\ \eta ^{klm}\
1,\; 
\]
whereas the choice of the solution (b) leads to the relation

\[
Q^kQ^lQ^m+Q^lQ^mQ^k+Q^mQ^kQ^l=0; 
\]
If $(k,l,m,n)$ is a set of indices such that $Q^kQ^lQ^mQ^n\not =0$, then the
following condition of consistency of representations must be satisfied

\[
\varrho _{klm}\,\varrho _{mkn}\,\varrho _{lkn}\,\varrho _{nlm}=1. 
\]
In this section we shall construct and explore some examples of the above
algebra with the tensor

\[
\eta ^{klm}=\pm {\frac 13}(1-\varrho _{klm}^2(\sigma
))^{-1}\sum_{p=1}^N\delta _p^k\delta _p^l\delta _p^m. 
\]
The above tensor satisfies the equation (a) and it leads to the algebras
whose generators are subjected to the following relations: $%
(Q^1)^3=(Q^2)^3=\ldots =(Q^N)^3=\pm 1$ and $Q^kQ^lQ^m=\varrho _{klm}(\sigma
)Q^lQ^mQ^k$ if there are at least two different indices in the triple $%
(k,l,m)$.

In the simplest case $N=1$ the above described construction leads us to the
algebras generated by $Q_{\pm }$ which satisfies the relation $Q_{\pm
}^3=\pm 1$. We shall denote the corresponding algebras by ${\cal C}_{\pm }^1$%
. As a vector space each of the algebras ${\cal C}_{\pm }^1$ is spanned by
the monomials $1,Q_{\pm },Q_{\pm }^2$. The operators $Q_{+}=\partial
_{\theta ^2}+\theta $ and $Q_{-}=j\partial _\theta +\theta ^2$ (where $%
\partial _\theta $ is the Grassmann derivation introduced in \ref{grassderiv}
and $\partial _{\theta ^2}$ gives $1$ when applied to $\theta ^2$ and $0$
when applied to $1$ or $\theta $) acting on the one-dimensional ${\Bbb Z}_3$%
-graded Grassmann algebra with the generator $\theta $ give the matrix
representation of ${\cal C}_{+}^1$ and ${\cal C}_{-}^1$.

In the case $N=2$ there are two representations $\varrho_{211}$, $%
\varrho_{122}$ and there is only one consistency condition to be satisfied

\[
(\varrho _{211}\,\varrho _{122})^2=1. 
\]
The above condition shows that the representations $\varrho _{211}$, $%
\varrho _{122}$ should be conjugate to each other, i.e. $\varrho _{211}=\bar
\varrho _{122}$. Let us choose $\varrho _{122}(\sigma )=j,\varrho
_{211}(\sigma )=j^2$. This choice of representations leads to the following
ternary relations

\[
Q^2Q^1Q^1=j^2\,Q^1Q^1Q^2=j\,Q^1Q^2Q^1\text{\quad and\quad }%
Q^1Q^2Q^2=j\,Q^2Q^2Q^1=j^2\,Q^2Q^1Q^2. 
\]
Let us rewrite the above relations in the following form:

\[
Q^k(Q^2Q^1-j\,Q^1Q^2)=0\text{\quad and\quad }(Q^2Q^1-j\,Q^1Q^2)Q^k=0\text{%
\quad with }k=1,2. 
\]
This form suggests that the ternary relations can be derived from the binary
ones $Q^2Q^1=j\,Q^1Q^2$. If one associates grade $1$ to the generators $%
Q^1,Q^2$ then the above binary relation implies that $Q^1,Q^2$ are ${\Bbb Z}%
_3$-commutative generators, i.e. $[Q^2,Q^1]_{{\Bbb Z}_3}=0$. Thus the
algebra generated by the ${\Bbb Z}_3$-commutative generators $Q^1,Q^2$ of
grade $1$ such that $(Q^1)^3=(Q^2)^3=1$ is a ${\Bbb Z}_3$-graded Clifford
algebra and it provides a realization of the general structure described at
the beginning of this section for $N=2$. Let us denote this algebra by $%
{\cal C}_{{\Bbb Z}_3}^2$.

Let us assume that ${\cal C}^1(Q^1)$ and ${\cal C}^1(Q^2)$ are two copies of
the one-dimensional ternary Clifford algebra generated respectively by the
elements $Q^1$ and $Q^2$. We define the ${\Bbb Z}_3$-{\it graded tensor
product} of the algebras ${\cal C}^1(Q^1)$ and ${\cal C}^1(Q^2)$ as the
tensor product of the underlying ${\Bbb Z}_3$-graded vector spaces endowed
with the multiplication defined as follows:

\[
(A\otimes B)(A^{\prime }\otimes B^{\prime })=j^{gr(B)gr(A^{\prime
})}AA^{\prime }\otimes BB^{\prime }, 
\]
where $A,A^{\prime }\in {\cal C}^1(Q^1)$ and $B,B^{\prime }\in {\cal C}%
^1(Q^2)$. Then ${\cal C}_{{\Bbb Z}_3}^2={\cal C}^1(Q^2)\otimes _{{\Bbb Z}_3}%
{\cal C}^1(Q^1)$.

Now we turn to the matrix representation of the algebra ${\cal C}_{{\Bbb Z}%
_3}^2$. The matrix algebras generated by a pair of $n\times n$ matrices $A$
and $B$ satisfying relations like $A^n=B^n=\pm 1$ and $AB=\mu BA$, for $\mu $
a primitive $n^{\text{th}}$ root of unity were studied by Sylvester \cite
{syl} in relation with the quaternion-like algebras. If $n=3$ the pair of
matrices $A,B$ gives the matrix realization of the ${\Bbb Z}_3$-graded
Clifford algebra ${\cal C}_{{\Bbb Z}_3}^2$. Sylvester called the elements of
this nine-dimensional matrix algebra {\it nonions}. The matrices
representing the algebra ${\cal C}_{{\Bbb Z}_3}^2$ are the following ones

\[
Q^1=\left( 
\begin{array}{ccc}
0 & 0 & 1 \\ 
1 & 0 & 0 \\ 
0 & 1 & 0
\end{array}
\right) \text{ and }Q^2=\left( 
\begin{array}{ccc}
0 & 0 & 1 \\ 
j & 0 & 0 \\ 
0 & j^2 & 0
\end{array}
\right) 
\]
where the corresponding operators acting on the Grassmann algebra ${\cal T}%
^1 $ have the form $Q^1=\partial _{\theta ^2}+\theta $, $Q^2=\partial
_{\theta ^2}+j\,\theta +(j^2-j)\,\theta ^2\partial _\theta $. This is made
clear by representing an element of the Grassmann algebra with the vector of
its components along $1$, $\theta $ and $\theta ^2$. So, the concept of a
ternary Clifford algebra throws light upon nonions from a new point of view
and leads to interesting conclusions.

\section{Ternary analogue of Orthogonal Group}

It is well known that Clifford algebras and their matrix representations
provide an appropriate framework for the spinor groups. Therefore the
natural question is whether there is an analogue of spinor groups covering
the group of orthogonal matrices in the case of the ${\Bbb Z}_3$-graded
Clifford algebra ${\cal C}_{{\Bbb Z}_3}^2$. It turns out that the answer to
this question is positive and the corresponding construction leads to the
interesting ternary analogue of orthogonal matrices.

We shall use the ternary forms $\eta $ and $\bar \eta $ with coefficients
transforming according to the laws

\[
\eta _{ijk}=j\,\eta _{jki}=j^2\,\eta _{kij}\text{\quad and\quad }\bar \eta
_{ijk}=j^2\,\bar \eta _{jki}=j\,\bar \eta _{kij} 
\]
as analogues of the skew-symmetric matrices. We shall call forms like $\eta $
the $j$-skew-symmetric forms and forms like $\bar \eta $ the $j^2$%
-skew-symmetric forms. Let us denote by ${\cal F}$ the vector space of the $%
j $-skew-symmetric forms and by $\bar {{\cal F}}$ the vector space of the $%
j^2$-skew-symmetric forms. It is clear that each pair of indices $(i,j)$, $%
\,i\not =j$ determines two linearly independent components $\eta _{ijj}$, $%
\eta _{jii}$ of the form $\eta $. Let us note $\eta _{211}=\omega $, $\,\eta
_{122}=\chi $. Then $\eta _{112}=j^2\,\omega $, $\,\eta _{121}=j\,\omega $
and $\eta _{212}=j\,\chi $, $\,\eta _{221}=j^2\,\chi $. It is useful in what
follows to associate to the form $\eta $ two vectors and two associated
complex numbers $\omega $ and $\chi $:

\[
\eta _v=(\eta _{121},\eta _{112},\eta _{211})=\omega .(j,j^2,1)\text{\quad
and\quad }{\eta }_v^{\prime }=(\eta _{212},\eta _{221},\eta _{122})=\chi
\,.(j,j^2,1) 
\]
and to consider the formal vector $Q_v=(Q^1,Q^2,Q^3)$ (where $%
Q^3=(Q^1)^2(Q^2)^2$) whose components are the grade $1$ monomials of the $%
{\Bbb Z}_3$-graded Clifford algebra ${\cal C}_{{\Bbb Z}_3}^2$. Let us define 
$\eta (Q_v)=\omega \,Q^2Q^1Q^1$ and $\,\bar \eta (Q_v)=\chi \,Q^1Q^2Q^2$.

Let us now define the map $m:{\bf C}^3\to ({\cal C}_{{\Bbb Z}_3}^2)_{\bar 1}$
by the formula

\[
m(z)=z_1\,Q^1+z_2\,Q^2+jz_3\,Q^3\text{\quad with }z=(z_1,z_2,z_3)\in {\bf C}%
^3 
\]
and the linear map $R$ from the vector spaces ${\cal F}$ and ${\bar {{\cal F}%
}}$ to the subalgebra $({\cal C}_{{\Bbb Z}_3}^2)_{\bar 0}$ by the formulae

\[
R(\eta )=\eta (Q_v)\text{\quad and\quad }R(\bar \eta )=\bar \eta (Q_v). 
\]
The following identities are easily verified

\begin{equation}
\lbrack R(\eta ),m(z)]_{{\Bbb Z}_3}=(j-1)\,m(\psi (\eta _v)z),  \label{clif1}
\end{equation}
\begin{equation}
\lbrack R(\bar \eta ),m(z)]_{{\Bbb Z}_3}=(j-1)\,m(\psi ({\eta ^{\prime }}%
_v)z).  \label{clif2}
\end{equation}
where the two mappings $\psi $ and $\varphi $ are defined by the formulae

\[
\psi (x,y,z)=\left( 
\begin{array}{ccc}
0 & 0 & x \\ 
y & 0 & 0 \\ 
0 & z & 0
\end{array}
\right) \text{\quad and\quad }\varphi (x,y,z)=\left( 
\begin{array}{ccc}
0 & y & 0 \\ 
0 & 0 & z \\ 
x & 0 & 0
\end{array}
\right) . 
\]
In analogy with the classical case we consider the subalgebra $({\cal C}_{%
{\Bbb Z}_3}^2)_{\bar 0}$ of the elements of grade $0$ of the algebra ${\cal C%
}_{{\Bbb Z}_3}^2$. It is spanned by the monomials $1$, $Q^2Q^1Q^1$ and $%
Q^1Q^2Q^2$. Let us introduce the notations $T=Q^2Q^1Q^1$ and $S=Q^1Q^2Q^2$.
The structure of the subalgebra $({\cal C}_{{\Bbb Z}_3}^2)_{\bar 0}$ is
completely described by the formulae

\[
T^{3k}=1\text{,\quad }T^{3k+1}=T\text{,\quad }T^{3k+2}=S\text{,}\quad
S^{3k}=1\text{,\quad }S^{3k+1}=S\text{,\quad }S^{3k+2}=T 
\]

Let us consider the elements of the subalgebra $({\cal C}_{{\Bbb Z}%
_3}^2)_{\bar 0}$ which have the form of Gaussian expressions

\[
e^{R(\eta )}=e^{\eta (Q_v)} 
\]
Let us denote by ${\cal S}$ the group generated by the above elements. This
group is commutative and any of its elements can be expressed as follows

\[
g(\omega )=\Theta (\omega )+\Psi (\omega )\,T+\Phi (\omega )\,S 
\]
where the coefficients are the functions

\[
\Theta (\omega )={\frac 13}(e^\omega +e^{j\omega }+e^{j^2\omega })\text{,}%
\quad \Psi (\omega )={\frac 13}(e^\omega +j\,e^{j^2\omega }+j^2\,e^{j\omega
})\text{,}\quad \Phi (\omega )={\frac 13}(e^\omega +j\,e^{j\omega
}+j^2\,e^{j^2\omega }) 
\]

If $g(\omega )$ and $g(\omega )^{\prime }$ are two elements of ${\cal S}$
then

\[
g(\omega )g(\omega ^{\prime })=g(\omega +\omega ^{\prime }). 
\]

It turns out that just as in the case of the classical Clifford algebra the
identities (\ref{clif1}) and (\ref{clif2}) can be realized on the group
level and we shall construct this realization only for the first one since
the realization of the second one can be done similarly.

Let us consider the functional $3\times 3$-matrices depending on complex
variable~$\omega $

\[
A(\omega )=\left( 
\begin{array}{ccc}
\alpha (\omega ) & \beta (\omega ) & \lambda (\omega ) \\ 
j^2\lambda (\omega ) & \alpha (\omega ) & j\beta (\omega ) \\ 
j^2\beta (\omega ) & j\lambda (\omega ) & \alpha (\omega )
\end{array}
\right) 
\]
where the entries are the functions

\[
\alpha (\omega )={\frac 13}(e^\omega e^{-j^2\omega }+e^{j\omega }e^{-\omega
}+e^{j^2\omega }e^{-j\omega }), 
\]
\[
\beta (\omega )={\frac 13}(e^\omega e^{-j^2\omega }+j\,e^{j\omega
}e^{-\omega }+j^2\,e^{j^2\omega }e^{-j\omega }), 
\]
\[
\lambda (\omega )={\frac 13}(j\,e^\omega e^{-j^2\omega }+e^{j\omega
}e^{-\omega }+j^2\,e^{j^2\omega }e^{-j\omega }). 
\]
Straightforward computations show that

\[
A(\omega )A(\omega ^{\prime })=A(\omega +\omega ^{\prime })\text{\quad
and\quad }\det A(\omega )=1. 
\]
Consequently the above one-parameter matrices form the commutative group we
shall denote by ${\cal S}^{*}$. We let $\pi :{\cal S}\to {\cal S}^{*}$ be
the map

\[
\pi (g(\omega ))=A(\omega ), 
\]
where $g(\omega )$ is an element of ${\cal S}$. One can see that $\pi $ is a
homomorphism of groups.

It can be proved that

\[
g^{-1}(\omega )\,m(z)\,g(\omega )=m(\pi (g(\omega ))z), 
\]
and the above formula gives the realization of the formula \ref{clif1} on
the group level.

Let $A$ be an arbitrary $3\times 3$-matrix

\[
A=\left( 
\begin{array}{ccc}
{\alpha } & {\beta ^{*}} & {\tilde \lambda } \\ 
{\tilde \alpha _1} & {\beta _1} & {\lambda _1^{*}} \\ 
{\alpha _2^{*}} & {\tilde \beta _2} & {\lambda _2}
\end{array}
\right) . 
\]

We define the {\it cyclic transposition} $t_c$ of the $3\times 3$-matrix $A$
as follows: the entries marked with asterisk undergo the counterclockwise
cyclic permutation and the entries marked with tilde undergo the clockwise
cyclic permutation. Applying this definition to the matrix $A$ one gets

\[
A^{t_c}=\left( 
\begin{array}{ccc}
{\alpha } & {\lambda _1^{*}} & {\tilde \alpha _1} \\ 
{\tilde \beta _2} & {\beta _1} & {\alpha _2^{*}} \\ 
{\beta ^{*}} & {\tilde \lambda } & {\lambda _2}
\end{array}
\right) . 
\]
Note that the cyclic transposition does not change the trace of a matrix and 
$A^{t_c^3}=A$, where $t_c^3$ means the cyclic transposition applied three
times. Then it can be shown that the one-parameter matrices $A(\omega )$
satisfy the condition

\[
A(\omega )\,A^{t_c}(\omega )\,A^{t_c^2}(\omega )={\bf Id}. 
\]
The above stated condition can be considered as the ternary generalization
of the classical orthogonality and the group ${\cal S}^{*}$ can be
considered as the analogue of the group of orthogonal matrices. Taking the
matrix $A$ in the exponential form $A=e^B={\bf Id}+B+{\frac 1{{2!}}}%
B^2+\ldots $ one can get the infinitesimal form of the ternary
orthogonality, that is

\[
B+B^{t_c}+B^{t_c^2}=0. 
\]
Note that the matrices $\psi (\eta _v)$ associated with a $j$-skew-symmetric
form $\eta $ satisfy above condition.

The matrix representation of the algebra ${\cal C}_{{\Bbb Z}_3}^2$ allows to
establish the analogue of the special case of the Mathai-Quillen formula 
\cite{quill}. Let $A$ be a skew-symmetric $2\times 2$-matrix

\[
A=\left( 
\begin{array}{cc}
0 & a \\ 
-a & 0
\end{array}
\right) . 
\]
Then the Mathai-Quillen formula in the case of the algebra ${\cal C}^2$ can
be written as follows:

\[
\limfunc{Str}(e^{{\frac 12}\gamma ^tA\gamma })=2i({\frac{\sinh \,ia}{ia}})\,%
\sqrt{\det A}. 
\]

The grade $0$ monomials $T,S$ of the algebra ${\cal C}_{{\Bbb Z}_3}^2$ have
the following matrix representation :

\[
T=Q^2Q^1Q^1=\left( 
\begin{array}{ccc}
1 & 0 & 0 \\ 
0 & j & 0 \\ 
0 & 0 & j^2
\end{array}
\right) \text{\quad and\quad }S=Q^1Q^2Q^2=\left( 
\begin{array}{ccc}
1 & 0 & 0 \\ 
0 & j^2 & 0 \\ 
0 & 0 & j
\end{array}
\right) . 
\]
Note that these matrices determine two different ${\Bbb Z}_3$-structures on
the complex space ${\bf C}^3$. Let us choose the ${\Bbb Z}_3$-structure
defined by the matrix $T$. We replace the supertrace used in the
Mathai-Quillen formula by the notion of the ${\Bbb Z}_3$-graded trace
(hypertrace) defined in section \ref{z3trace}. Now the analogue of the
Mathai-Quillen formula has the form

\[
\limfunc{tr}\nolimits_{{\Bbb Z}_3}(e^{\eta (P_v)})=3\omega ^{-1}\,\Phi
(\omega )\sqrt[3]{\det {(\psi (\eta _v))}} 
\]

For a more detailed exposition, see \cite{abramovagg}.

\section{${\Bbb Z}_3$-analogue of the Dirac equation}

The ${\Bbb Z}_3$-graded generalization of Grassmann algebra discussed in
section \ref{z3grass} leads to a natural generalization of the superfields
as the fields composed of different contributions proportional to all
possible monomials in the ${\Bbb Z}_3$-graded generators $\theta ^A$ and $%
\bar \theta ^B$, i.e. 
\[
\Phi (\theta ^A,\bar \theta ^B,x^\mu )=\phi _0(x^\mu )+\psi _A(x^\mu )\theta
^A+\bar \theta ^B\bar \psi _B+\chi _{AB}\theta ^A\theta ^B+\ldots 
\]

However, this approach is made in the context of the {\em second quantization%
}, with operator-valued fields. One could ask if the ternary character of
such a theory could not be perceived even at a deeper level, i.e. in the
algebraic properties of the complex valued wave functions which would be the
solutions of some Schr\"odinger-like differential equations of a new type.

The usual (binary) Clifford algebra appears in a natural manner in Dirac's
equation which is, in a sense, a ``square root'' of the Klein-Gordon
equation. With the use of the ternary Clifford algebra defined above, the $%
{\Bbb Z}_3$-graded generalization of Dirac's equation should read: 
\[
\frac{\partial \,\psi }{\partial \,t}=Q^1\frac{\partial \psi }{\partial \,x}%
+Q^2\frac{\partial \psi }{\partial y}+Q^3\frac{\partial \psi }{\partial z}%
+Tm\psi 
\]
where $\psi $ stays for a triplet of wave functions, which can be considered
either as a column, or as a grade $1$ matrix with three non-vanishing
entries $u$ $v$ $w$, 
\[
Q^1=\left( 
\begin{array}{ccc}
0 & 0 & 1 \\ 
1 & 0 & 0 \\ 
0 & 1 & 0
\end{array}
\right) ,Q^2=\left( 
\begin{array}{ccc}
0 & 0 & 1 \\ 
j & 0 & 0 \\ 
0 & j^2 & 0
\end{array}
\right) ,Q^3=\left( 
\begin{array}{ccc}
0 & 0 & 1 \\ 
j^2 & 0 & 0 \\ 
0 & j & 0
\end{array}
\right) 
\]
and $T$ is the diagonal $3\times 3$ matrix with the eigenvalues $1$, $\,j$
and $j^2$. It is interesting to note that this is possible only with {\em %
three} spatial coordinates.

In order to diagonalize this equation, we must act three times with the same
operator, which will lead to the equation of {\em third} order, satisfied by
each of the three components $u$, $\,v$, $\,w$, e.g.: 
\[
\frac{\partial ^3\,u}{\partial t^3}=\left[ \frac{\partial ^3}{\partial x^3}+%
\frac{\partial ^3}{\partial \,y^3}+\frac{\partial ^3}{\partial z^3}-3\frac{%
\partial ^3}{\partial x\partial y\partial z}\right] \,u+m^3\,u 
\]

This equation can be solved by separation of variables; the time-dependent
and the space-dependent factors are linear combinations of $\Theta (\omega
t) $, $\Psi (\omega t)$, $\Phi (\omega t)$ and $\Theta ({\bf k}.{\bf r})$, $%
\Psi ({\bf k}.{\bf r})$, $\Phi ({\bf k}.{\bf r})$. Their nine independent
products can be represented in a basis of real functions as 
\[
\left( 
\begin{array}{ccc}
A_{11}\,e^{\omega \,t+{\bf k}.{\bf r}} & A_{12}\,e^{\omega \,t-\frac{{\bf k}.%
{\bf r}}2}\,cos\,\xi & A_{13}\,e^{\omega \,t-\frac{{\bf k}.{\bf r}}%
2}\,sin\,\xi \\ 
A_{21}\,e^{-\frac{\omega \,t}2+{\bf k}.{\bf r}}\,cos\,\tau & A_{22}\,e^{-%
\frac{\omega \,t}2-\frac{{\bf k}.{\bf r}}2}\,cos\,\tau \,cos\,\xi & 
A_{23}\,e^{-\frac{\omega \,t}2-\frac{{\bf k}.{\bf r}}2}\,cos\,\tau \sin \,\xi
\\ 
A_{31}\,e^{-\frac{\omega \,t}2+{\bf k}.{\bf r}}\,sin\,\tau & A_{32}\,e^{-%
\frac{\omega \,t}2-\frac{{\bf k}.{\bf r}}2}\,sin\,\tau \,cos\,\xi & 
A_{33}\,e^{-\frac{\omega \,t}2-\frac{{\bf k}.{\bf r}}2}\,sin\,\tau \,sin\,\xi
\end{array}
\right) 
\]
where $\tau =\frac{\sqrt{3}}2\omega \,t$ and $\xi =\frac{\sqrt{3}}2{\bf k}.%
{\bf r}$.

The parameters $\omega $, ${\bf k}$ and $m$ must satisfy the cubic
dispersion relation: $\omega ^3=k_x^3+k_y^3+k_z^3-3\,k_xk_yk_z+m^3$.
Although neither of the solutions belong to the space of tempered
distributions, it is possible to combine them into solutions of the ordinary
Klein-Gordon equation: The ternary skew-symmetric products contain only
trigonometric functions, depending on the combinations $2\,(\tau -\xi )$ and 
$2\,(\tau +\xi )$. As a matter of fact, not only the {\em determinant}, but
also each of the {\em minors} of this matrix is a combination of the
trigonometric functions only. The same is true for the binary products of
``conjugate'' solutions, with opposite signs for $\omega t$ and ${\bf k.r}$
in the exponentials. It is possible to find new parameters, which are linear
combinations of $\omega $, ${\bf k}$ and $m$, that will satisfy quadratic
relations that may be intrpreted as a mass shell equation.

\section{Conclusion}

It is clear from these examples that most of the mathematical structures
that are commonly used in supersymmetric theories can be generalized from
the ${\Bbb Z}_2$-graded case to the ${\Bbb Z}_3$-graded case. These
generalizations, though, are very different in their spirit from other
generalizations known as fractional supersymmetry.

The most important difference is that {\em ternary} rather than {\em binary}
relations define the algebraic structures of functions and fields.

Moreover, the ternary principle is extended in a natural way to all the
algebraic structures used in the classical field theories, such as
Grassmann, Lie and Clifford algebras.

The obvious inspiration for investigating this type of mathematical
structures comes from the idea that the confinement of quarks should have
its origin in the special algebraic structure of the corresponding fields,
which might elude the usual rules of quantum field theory, as well as the
observation in the form of a free field. We hope that further work in this
direction will contribute to shred some light on these problems.

\vspace{0.5cm}

{\bf Acknowledgments}

V. Abramov wishes to express his thanks to the {\it TEMPUS} foundation that
supported his 3-month stay in the {\it Laboratoire de Gravitation et
Cosmologie Relativistes} of the {\it Universit\'e Pierre et Marie Curie}.

B. Le Roy acknowledges the support of the {\it Minist\`ere de l'Enseignement
Sup\'erieur et de la Recherche}

\end{document}